\documentclass{article}
\usepackage{arxiv}

\usepackage[utf8]{inputenc} 
\usepackage[T1]{fontenc}    
\usepackage{hyperref}       
\usepackage{url}            
\usepackage{booktabs}       
\usepackage{amsfonts}       
\usepackage{nicefrac}       
\usepackage{microtype}      
\usepackage{amsmath,amssymb,amsfonts}
\usepackage{multirow}
\usepackage{graphicx}

\title{Unsupervised Interpretable Representation Learning for Singing Voice Separation}

\author{
  Stylianos I. Mimilakis \\
  Semantic Music Techn. Group\\
  Fraunhofer-IDMT\\
  Ilmenau, Germany\\
  \texttt{mis@idmt.fraunhofer.de} \\
   \And
 Konstantinos Drossos \\
  Audio Research Group\\
  Tampere University\\
  Tampere, Finland \\
  \texttt{konstantinos.drossos@tuni.fi} \\
     \And
 Gerald Schuller \\
  Applied Media Systems Group\\
  Technical University of Ilmenau\\
  Ilmenau, Germany \\
  \texttt{gerald.schuller@tu-ilmenau.de} \\
}

\begin{document}
\maketitle

\begin{abstract}
In this work, we present a method for learning interpretable music signal representations directly from waveform signals. Our method can be trained using unsupervised objectives and relies on the denoising auto-encoder model that uses a simple sinusoidal model as decoding functions to reconstruct the singing voice. To demonstrate the benefits of our method, we employ the obtained representations to the task of \textit{informed} singing voice separation via binary masking, and measure the obtained separation quality by means of scale-invariant signal to distortion ratio. Our findings suggest that our method is capable of learning meaningful representations for singing voice separation, while preserving conveniences of the the short-time Fourier transform like non-negativity, smoothness, and reconstruction subject to time-frequency masking, that are desired in audio and music source separation.
\end{abstract}

\keywords{Representation learning, unsupervised learning, denoising auto-encoders, singing voice separation} 

\section{Introduction}
A particular task in music signal processing that has attracted a lot of research interest is the estimation of the singing voice source from within an observed mixture signal~\cite{rafii18}. To that aim, deep supervised learning is shown to yield remarkable results. Approaches that rely on deep supervised learning can be discriminated in two categories, the ones that operate in the short-time Fourier transform (STFT) domain~\cite{stoter19, spleeter2019}, and we denote as spectral-based approaches, and the ones that operate directly on the waveform signals~\cite{demucs, meta_tasnet}, that we denote as waveform-based approaches. Spectral and waveform based approaches 
have in common that they implicitly compute source-dependent masks that are applied to the mixture signal, prior to the reconstruction of the target signals~\cite{stoter19, spleeter2019, demucs, meta_tasnet}\footnote{Regarding the masking strategy, we are referring to the adaptation of Conv-TasNet for music signals also presented in~\cite{demucs}.}. 

Although the implicit masking is shown to be a simple and robust method to learn source dependent patterns for source separation~\cite{mappings2020}, one could expect that waveform based approaches would significantly outperform the spectral ones. That is because waveform based approaches are optimized using time-domain signals that also contain the phase information, that unarguably carries important signal information~\cite{magon_misi,magron:2018:interspeech} and has been neglected by many spectral based approaches~\cite{stoter19, spleeter2019, drossos18, mim18}. However, experimental evidence shows that spectral based approaches have comparable or marginally better separation performance to the waveform ones~\cite{demucs, meta_tasnet, stoter19}. Since the state-of-the-art (SOTA) methods for both waveform and spectral approaches rely on deep neural networks, and in both spectral and waveform approaches a considerable engineering effort has been directed to the employed neural architecture, it is evident that the difference in the performance between the two different approaches can be attributed to the utilized signal representation. For the waveform-based ones this is the output of an encoder
, but for the spectral-based it is the non-negative signal representation offered by the magnitude of the STFT. Thus, we believe that learning generalized signal representations for music signals is an intriguing direction for music source separation research.

In this work, we focus on representation learning~\cite{simply_bengio_rl} for singing voice separation in an attempt to bridge the gap between spectral and waveform based approaches. To this aim, we propose a simple method for unsupervised representation learning from waveform signals, alleviating the need of having paired training data (i.e., matched multi-track audio data). However, the method still requires isolated source's audio signals. More specifically, our method is based on the denoising auto-encoder (DAE) model~\cite{vincent_den}, but for the decoding functions our model for representation learning inherits a simple and real-valued sinusoidal model. The sinusoidal model consists of amplitude-modulated cosine functions, and whose parameters are jointly optimized with the rest of the DAE. The motivation behind using a sinusoidal model as a decoding function is to guide (via back-propagation) the encoding layers to learn and convey information regarding the energy of specific cosine functions that compose the audio signal, leading to interpretable representations. 

Our method is inspired by the concept of differentiable digital signal processing~\cite{ddsp} where the parameters of common digital signal processing functions are optimized by means of back-propagation, and in our case we back-propagate through the parameters of a simple signal model. Furthermore, our method is similar to the Sinc-Network presented in~\cite{sinc_net}, that uses sinc functions in the encoding layers of convolutional kernels for interpretable deep learning, and its extension to complex-valued representations for speaker separation~\cite{filterbank_design_e2e}.
However, our method differs from~\cite{filterbank_design_e2e} as the representation of the proposed method is real-valued, alleviating the cumbersome signal processing operations on complex numbers. It also differs from approaches that initialize the front-end parts of the networks with cosine functions~\cite{adaptive_fe_ss} that are then updated by means of back-propagation, by inheriting the cosine functions as a part of the model to be optimized. Finally, our method provides an unsupervised alternative to the source informed method for representation learning presented in~\cite{tzinistwostep}. The rest of the document is organized as follows: Section~\ref{sec:proposed_model} presents the proposed method, Section~\ref{sec:experiments} describes the followed experimental procedure, Section~\ref{sec:results} discusses the obtained results, and Section~\ref{sec:conclusions} concludes this work.

\section{Proposed Method}\label{sec:proposed_model}
\begin{figure}[!t]
    \centering
    \includegraphics[width=0.7\columnwidth, keepaspectratio]{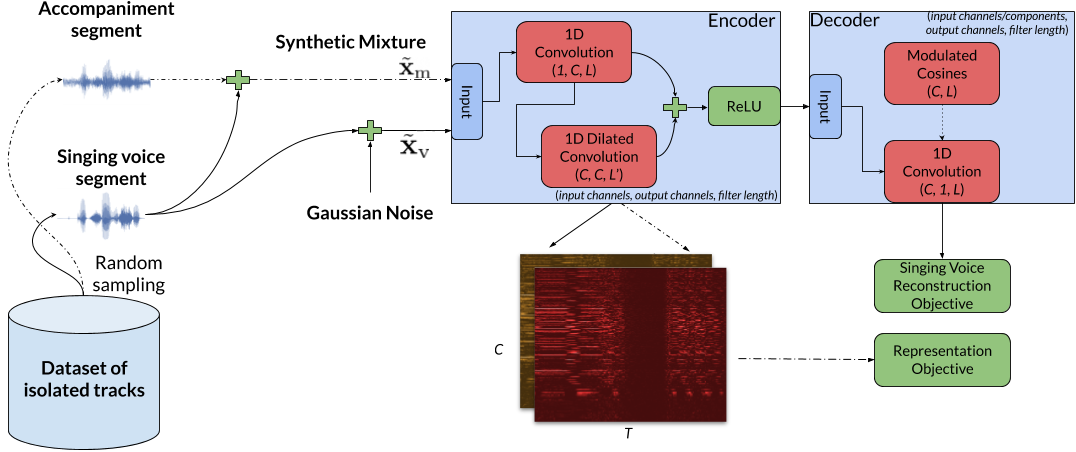}
    \caption{Overview of the proposed method.}
    \label{fig:ov-method}
\end{figure}

Our proposed method employs two functions, the encoder $E$, and the decoder $D$. The input to our method is a music signal $\mathbf{x} \in \mathbb{R}^{N}$ of $N$ time-domain samples, and the output is the learned representation of $\mathbf{x}$, denoted as $\mathbf{A} \in \mathbb{R}^{C\times T}$. $C$ is the number of templates and $T$ is the temporal length of each template (similarly as the time-frames in STFT-related representations). The encoder $E$ learns the representation $\mathbf{A}$ with the help of the decoder $D$. $D$ is responsible for reconstructing the signal, given the representation computed by $E$. The reconstructed signal can then be used to optimize $E$ and $D$ using a reconstruction objective. To enforce interpretability for the representation $\mathbf{A}$, we use a differentiable sinusoidal synthesis model for the decoder $D$. An illustration of the proposed method is given in Figure~\ref{fig:ov-method}.

\subsection{The Encoder}
During inference, the encoder $E$ gets as an input any music signal $\mathbf{x}$ and outputs its representation $\mathbf{A}$. In order for $E$ to yield the representation $\mathbf{A}$, an initial stage of training is performed. During training, two synthetic signals are used. Each synthetic signal employs the singing voice signal ($\mathbf{x}_{\text{v}} \in \mathbb{R}^{N}$). The first synthetic signal is termed as $\tilde{\mathbf{x}}_{\text{m}} \in \mathbb{R}^{N}$ and is the result of a corruption process for $\mathbf{x}_{\text{v}}$ with an additive generic multi-modal distribution-based noise (e.g. a randomly selected signal that contains accompaniment music, like a mixture of drums, guitars, synthesizers, and bass). The second signal is termed as $\tilde{\mathbf{x}}_{\text{v}}$ and is the result of a corruption process for $\mathbf{x}_{\text{v}}$ using additive Gaussian noise. 

Both signals $\tilde{\mathbf{x}}_{\text{m}},\, \tilde{\mathbf{x}}_{\text{v}}$ are used independently as an input to $E$, resulting into two representations $\mathbf{A}_{\text{m}},\,{\mathbf{A}}_{\text{v}} \in \mathbb{R}^{C\times T}$, respectively. To compute each representation, $E$ consists of two one-dimensional strided convolutions, with appropriate zero-padding. The first operation involves a convolution of each signal with a set of $C$ number of kernels of temporal length $L$ and a stride $S$. The stride $S$ is a hyper-parameter and affects the expected number of time-frames $T$ by $T=\lceil N/S \rceil$, where $\lceil\cdot\rceil$ is the ceiling function. The resulting latent signal is given to the second convolution, which is a \textit{dilated} one-dimensional convolution~\cite{dilated_convs} with $C$ number of kernels, a smaller temporal length $L' << L$, and a stride equal to 1. The output of the second convolution is updated by means of residual connections using the output from the first convolution, followed by the rectified linear unit (ReLU) activation function~\cite{relu_rbm}. The ReLU function promotes a non-negative and sparse representation by preserving positive values and setting the rest to zero~\cite{papyan17}, and is shown to be particularly useful in general modelling of audio signals~\cite{non_negative_paris}.
 
Another targeted (and useful) property of the representation is that of smoothness~\cite{adaptive_fe_ss, non_negative_paris}, especially useful when real-valued cosine functions are involved in auto-encoding or separation models~\cite{adaptive_fe_ss}. That is because audio signal modelling based on cosine functions requires the phase information for reconstruction. Phase information is usually encoded as the sign (positive or negative value) of the real-valued representation that varies along the time-frames of the representation. Since the negative values are nullified by the application of the ReLU function, neighbouring time-frames, that convey similar information for music signals are expected to be non-smooth. To compensate for that, the second convolution operation of $E$ is using dilated convolutions that aggregate temporal information from neighboring time-frames~\cite{dilated_convs,drossos:2020:ijcnn}.

In order to enforce the learning of smooth representations, we employ a representation objective that the encoder has to minimize. Specifically, we use the representation of $\tilde{\mathbf{x}}_{m}$, ${\mathbf{A}}_{\text{m}}$, to compute the total variation denoising~\cite{tv_loss} ($\mathcal{L}_{\text{TV}}$) as
\begin{align}
    \label{eq:tv_loss}
    {\mathcal{L}_{\text{TV}}}(\mathbf{A}_\text{m}) &= \frac{1}{CT} \Big( \sum_{c=1}^{C-1}||{\mathbf{A}_\text{m}}_{[c;\ldots]} - {\mathbf{A}_\text{m}}_{[c-1;\ldots]}||\nonumber\\ 
    &+ \sum_{t=1}^{T-1} ||{\mathbf{A}_\text{m}}_{[\ldots; t]} - {\mathbf{A}_\text{m}}_{[\ldots; t-1]}||\Big)\text{ ,}
\end{align}
\noindent
where ${\mathbf{A}_\text{m}}_{[c;\ldots]}$ and ${\mathbf{A}_\text{m}}_{[\ldots; t]}$ are the $c$-th row and $t$-th column vectors (respectively) of the matrix $\mathbf{A}_{\text{m}}$, and $||\cdot||$ is the $\ell_1$ vector norm. Eq.~\eqref{eq:tv_loss} penalizes $E$ by the norm of the first order difference across both time-frames $T$ and templates $C$. The former promotes slow time varying representations as the magnitude of the STFT representation, and the latter promotes a grouping of the template activity. We use $\mathbf{A}_\text{m}$ only to compute $\mathcal{L}_{\text{TV}}$, to enforce the encoder $E$ to yield smooth representations on the most realistic corruption scenario. This scenario is the additive generic multi-modal distribution-based noise $\tilde{\mathbf{x}}_{\text{m}}$ that contains also the information regarding the singing voice signal $\mathbf{x}_{\text{v}}$. Thus, the smoothness for the representation of the singing voice is implicitly enforced.

\subsection{The Decoder}\label{sub-sec:decoder}
The decoder $D$ accepts the representation $\mathbf{A}_{\text{v}}$, and yields $\hat{\mathbf{x}}_{\text{v}}$ which is the approximation of the clean singing voice $\mathbf{x}_{\text{v}}$. Specifically, $D$ models $\mathbf{x}_{\text{v}}$ as a sum of $C$ signal components that overlap in $\mathbb{R}^{N}$. The components are computed by a strided convolution\footnote{Appropriate zero-padding is assumed to be applied in order to deal with the differences between $T$ and $L$.} between the representation template $\mathbf{A}_{\text{v}_{[c; \ldots]}}$ and the kernel $\mathbf{w}_{c} \in \mathbb{R}^{L}$ of temporal length $L$ as
\begin{equation}\label{eq:1}
    \mathbf{x}_{\text{v}} \approx \hat{\mathbf{x}}_{\text{v}} := \sum_{c=0}^{C-1}\mathbf{A}_{\text{v}_{[c; \ldots]}} * \mathbf{w}_{c} \text{.}
\end{equation}
\noindent

Similar to Sinc-Net~\cite{sinc_net} and it's complex-valued extension for speech enhancement~\cite{filterbank_design_e2e}, we do not allow each $\mathbf{w}_{\text{c}}$ to be updated directly using back-propagation. Instead, we re-parameterize each $\mathbf{w}_{\text{c}}$ using sinusoidal functions and back-propagate through their corresponding parameters. More specifically, we compute each $\mathbf{w}_{\text{c}}$ using
\begin{equation}
    \label{eq:2}
    \mathbf{w}_{c} = \text{cos}(2 \pi f_{c}^{2} \odot \mathbf{t} + \phi_{c}) \odot  \mathbf{b}_{c}\text{ ,}
\end{equation}
\noindent
where $\text{cos}$ and $\odot$ are the element-wise cosine function and product, respectively, and $\mathbf{t} \in \mathbb{Z}^{L}$ is a vector denoting the integer time indices $[0, \ldots, L-1 ]$ of the kernels. These parameters of the cosine function are considered constants and are shared between the kernels. The sampling-rate-normalized carrier frequency $f_{c}$, the phase $\phi_{c}$ (in radians), and the modulating signal $\mathbf{b}_{c}$ are learnable and different for each kernel. The non-linear squaring operation applied to $f_{c}$ is motivated by the increased frequency resolution in lower frequencies that music signals commonly have~\cite{rafii18}, and is an experimental finding that is studied in Section~\ref{sec:results}. 
Using Eq.~\eqref{eq:2} for all $C$, our method constructs $\mathbf{W} \in \mathbb{R}^{C \times L}$ by stacking the corresponding outcome.
After the stacking, a sorting operation is applied to $\mathbf{W}$, which sorts the kernels $\mathbf{w}_{c}$ in ascending order based on the normalized and squared carrier frequency $f_c$, promoting an intuitive representation. Then the decoding operation for $\mathbf{A}_\text{v}$ takes place using Eq.~\eqref{eq:1}.

There are three reasons for using modulated cosine functions for decoding $\mathbf{A}_{\text{v}}\,$: a) cosine functions promote interpretability~\cite{sinc_net}, i.e., the representation $\mathbf{A}$ is expected to convey amplitude related information for driving a well established synthesis model based on sinusoidal functions~\cite{serra_sms}, b) the auto-encoding operation shares many similarities with the STFT yet without having to deal directly with the phase information, for which supervised based separation works remarkably well~\cite{stoter19, spleeter2019}, and c) amplitude modulations allow an extra degree of freedom in reconstructing signals that cannot be described by pure sinusoidal functions~\cite{serra_sms}. The latter statement is supported by the convolution theorem which states that the element-wise product of two vectors can be expressed in the Fourier domain as their corresponding convolution. Since in our re-parameterization scheme (i.e., Eq.~\eqref{eq:2}) one of the signals is a cosine function, then $\mathbf{b}_{c}$ is expected to convey information regarding fricatives and/or formants of the singing voice signal $\mathbf{x}_{\text{v}}$. Regarding on whether the proposed decoder is efficient in reconstructing the singing voice compared to either cosine functions or commonly employed convolutional layers, the reader is kindly referred to Section~\ref{sec:results}.

The optimization objective for $D$ is the negative signal-to-noise ratio (neg-SNR)~\cite{uni_ass}, defined as:
\begin{equation}
    \label{eq:neg-snr}
    \mathcal{L}_{\text{neg-SNR}} (\mathbf{x}_{\text{v}}, \hat{\mathbf{x}}_{\text{v}}) = - 10 \, \text{log}_{10}\Big(\frac{||\mathbf{x}_{\text{v}}||_{2}^2}{||\mathbf{x}_{\text{v}} - \hat{\mathbf{x}}_{\text{v}}||_{2}^2}\Big) \text{ , where}
\end{equation}
\noindent
$||\cdot||_{2}^{2}$ is the squared $\ell_2$ vector norm, and the negative sign is used to cast the logarithmic SNR as a minimization problem. Using Eq.\eqref{eq:neg-snr} and Eq.\eqref{eq:tv_loss} the overall minimization objective is
\begin{equation}\label{eq:total_loss}
\mathcal{L} = \mathcal{L}_{\text{neg-SNR}} + \lambda\mathcal{L}_{\text{TV}}
\end{equation}
\noindent
where $\lambda$ is a scalar for weighting the impact of Eq.\eqref{eq:tv_loss} in the learning signal. The decoder $D$ computes $\hat{\mathbf{x}}_{\text{v}}$ only from the signing voice representation $\mathbf{A}_{\text{v}}$. That is because we aim at learning general representations in an unsupervised and not discriminative fashion. To achieve that by means of the DAE model~\cite{vincent_den}, we assume that the distribution of the corruption process is constant for all segments in the data-set~\cite{score_matching_daes}. This cannot be assumed for music signal mixtures, as even the distribution of the accompaniment instruments can vary dramatically from one segment to another. Consequently, by making such an assumption it could lead to degenerate representations for singing voice.

\section{Experimental Procedure}\label{sec:experiments}
For training and evaluating the proposed method we use the MUSDB18 data-set~\cite{musdb18} that consists of 150 two-channel multi-tracks, sampled at 44100Hz and split into training (100 multi-tracks) and testing (50 multi-tracks) subsets. During training we sample a set of four multi-tracks from which we use the vocals and the accompaniment sources. Each sampled multi-track is down-mixed to a single channel and is partitioned into overlapping segments of $N=44100$ samples with an overlap of 22050 samples. We then randomly shuffle the segments for each source and corrupt the singing voice signal as described in Section~\ref{sec:proposed_model}. The standard deviation of the additive Gaussian noise corruption is set to $1e-4$ and is independent from the signal's amplitude. A batch of 8 segments is used for optimizing the parameters of the proposed method, minimizing Eq.~\eqref{eq:total_loss} using adam algorithm~\cite{adam} with a learning rate of $1e-4$. For choosing the convolution hyper-parameters we conducted an informal experiment employing 20 tracks from the training subset,
followed by informal listening tests. This resulted into the following hyper-parameters: $C=800$, $S=256$, $L=2048$, $L'=5$,  $D=10$, and $\lambda = 0.5$. During the informal experiments, we observed that the method converges fast so we set the total number of iterations throughout the whole data to $10$. The choice for $N=44100$ samples was based on the available computational resources.

For evaluation we use the rest 50 tracks, that are down-mixed and partitioned into non-overlapping segments. The shuffling and random mixing are not considered in the evaluation stage, but silent segments are discarded. We test the usefulness of the representation by performing informed and masking-based singing voice separation, following the recently proposed framework for assessing latent representations for audio source separation~\cite{tzinistwostep}. To that aim, we employ the trained decoder (according to the previously described procedure), and reconstruct the time-domain signals of the un-corrupted singing voice representation and the binary masked mixture representation, respectively. The binary mask is computed using the encoded singing voice, accompaniment, and their corresponding mixture signals, that are available in the test sub-set.
The reconstructed time-domain signals are used for computing the scale-invariant signal-to-distortion ratio (SI-SDR)~\cite{si_sdr} defined as
\begin{equation}\label{eq:si-sdr}
    \text{SI-SDR}(\mathbf{x}_{\text{v}}, \hat{\mathbf{x}}_{\text{v}}) = 10\,\text{log}_{10} \Big(\frac{||\alpha \mathbf{x}_{\text{v}}||_{2}^2}{||\alpha\mathbf{x}_{\text{v}}-\hat{\mathbf{x}}_{\text{v}}||_{2}^2}\Big) \text{, for } \alpha = \frac{\tilde{\mathbf{x}}_{\text{v}}^{T}\mathbf{x}_{\text{v}}}{||\mathbf{x}_{\text{v}}||_{2}^{2}}\text{ ,}
\end{equation}
\noindent
and is used computed for each segment. In the following section, we report the median value of SI-SDR across segments and three experimental runs.

Using the above described procedure, we conduct two experiments. In the first experiment, we examine whether the modulated cosine functions (\texttt{mod-cos}) are a good synthesis model by measuring the reconstruction performance, after being optimized for the denoising task. We optimize various models that use the proposed training scheme presented in Section~\ref{sec:proposed_model} \textit{without} the random mixing corruption processes, and by employing the early stopping mechanism to terminate the training procedure if the model has stopped decreasing the loss expressed in Eq.~\eqref{eq:neg-snr} during the updates of the previous epoch. We consider various decoding strategies such as non-modulated cosine functions (\texttt{cos}), and common one-dimensional convolutional networks (\texttt{conv}) with and without the tanh non-linearity at the output. We also examine Sinc-Net~\cite{sinc_net} (\texttt{sinc}) as the first encoding stage as proposed in~\cite{sinc_net}. In this experiment $C$ is adapted so that each model uses approximately the same number of parameter.

For the second experiment we re-train the best combination of the above, using various values for the number of components $C \in [400, 800, 1600]$ and perform the reconstruction of the binary masked mixture signal. To examine the regularization effect of the total-variation (Eq.~\eqref{eq:tv_loss}) computed using the random mixing corruption process, we report each model's performance by using Eq.~\eqref{eq:tv_loss} for both $\mathbf{A}_{\text{v}}$ and $\mathbf{A}_{\text{m}}$, respectively. For comparison, we employ the STFT and perform the above described operations of analysis, masking, and synthesis. The STFT uses a hop-size of 384 samples, a window size of 2048 samples, and the hamming windowing function. The difference between the first and the second set of experiments is that for the second set of experiments the modulated cosine functions are sorted after each gradient update, as explained in Section~\ref{sub-sec:decoder}, whereas in the first they are not. The sorting is performed for the representation to have information analogous other cosine related transforms.

\section{Results \& Discussion}\label{sec:results}
The obtained results from the two experiments are presented in Tables~\ref{tab:res-1} and~\ref{tab:res-2}. Additional results, illustrations underlining the interpretability of the representations, and audio examples can be found online\footnote{\url{https://github.com/Js-Mim/rl_singing_voice}}. 
Table~\ref{tab:res-1} demonstrates the median SI-SDR expressed in dB (the higher the better) yielded by the first experiment, along with additional information regarding the various setups for the encoder $E$ and the decoder $D$, the number of parameters $N_P$ (in millions M), the used number of components $C$, and the employed non-linearities. The results in Table~\ref{tab:res-1} highlight three trends. First, the application of the non-linearity to the normalized frequencies $f_c$ results into better reconstruction performance compared to the linear case. The observed improvement is of $\sim5$dB on average across experimental configurations. Secondly, the modulated cosine functions serve as a good differentiable synthesis model for singing voice signals, outperforming simple cosine functions by approximately 8 dB on average, with respect to the two experimental configurations (with and without frequency scaling of the normalized frequency), and by $1.4$ dB the best configuration of convolution based model (\texttt{conv}). Since SI-SDR is invariant to scale modifications of the assessed signal, $1.4$ dB is a significant improvement of signal quality and does not imply a simple matching of the gain that the model based on modulated cosine functions might have exploited. Thirdly, Sinc-Net~\cite{sinc_net} does not bring further improvements to the proposed method.
\begin{table}[!ht]
\centering
\caption{Results reflecting the decoding performance, by means of SI-SDR. Bold-faced numbers denote the best performance.}
\resizebox{0.56\columnwidth}{!}{%
\begin{tabular}{c|c|c|c|c}  
$E/D$ Setup & Non-linearity  & $C$ & SI-SDR & $N_P$    \\ \hline
\multirow{2}{*}{\small{\texttt{conv/cos}}} & N/A & \multirow{2}{*}{952} & 20.83 & \multirow{2}{*}{6.483M} \\
 {} &  $f_{c}^{2}$ & & 22.34 & {} \\ \hline
\multirow{2}{*}{\small{\texttt{conv/conv}}} & N/A & \multirow{2}{*}{800} & 31.25 & \multirow{2}{*}{6.476M} \\
 {} & tanh(decoder) & {} & 30.50 & {} \\ \hline
 \multirow{2}{*}{\small{\texttt{conv/mod-cos}}} & N/A & \multirow{2}{*}{800} & 28.72 & \multirow{2}{*}{6.478M} \\
 {} & $f_{c}^{2}$ & {} & \bf{32.62} & {} \\ \hline
 {\small \texttt{sinc/mod-cos}} & $f_{c}^{2}$ & {952} & 26.82 & 6.487M \\ 
\end{tabular}}
\label{tab:res-1}
\end{table}

Focusing on the separation performance of the obtained representations, Table~\ref{tab:res-2} presents the median SI-SDR values of the binary masking separation scenario for three values for the hyper-parameter $C$ and two regularization strategies including two different signal representations, the corrupted by Gaussian noise $\mathbf{A}_{\text{v}}$, and the synthetic mixtures using the accompaniment signals $\mathbf{A}_{\text{m}}$. The obtained results are compared to the STFT that has perfect reconstruction properties and masking techniques work very well in practice~\cite{rafii18}. The results of Table~\ref{tab:res-2} underline two main experimental findings. The first finding is that the binary masking can be used to separate sources using the proposed approach for representation learning. This can be seen from the $C=1600$ model that uses the synthetic mixtures as an input to the unsupervised representation objective and achieves a SI-SDR median value of $6.68$ dB. The second finding is that the proposed unsupervised representation objective, i.e., Eq.~\eqref{eq:tv_loss} with the synthetic mixtures, can be used to improve the reconstruction of the masked mixture signals without additional supervision, as previous studies suggest~\cite{tzinistwostep}. This claim is supported by the observed improvement of $\sim 2$ dB, on average across models of various components $C$, when the synthetic mixtures are used for the unsupervised representation objective. Nonetheless, there is much room for improvements in 
order to obtain the quality of the STFT/iSTFT approach that outperforms the best masked approximation of the proposed method by $2.12$ dB.
\begin{table}[!th]
\centering
\caption{SI-SDR for informed separation by binary masking (BM). Bold-faced numbers denote the best performance.}
\resizebox{0.65\columnwidth}{!}{%
\begin{tabular}{c|c|c|c|c|c}  
$E/D$ Setup & $C$ & ${\mathcal{L}_{\text{TV}}}(*)$ & SI-SDR & BM SI-SDR& $N_P$    \\ \hline
\multirow{6}{*}{\small{\texttt{conv/mod-cos}}} & \multirow{2}{*}{400} & $\mathbf{A}_\text{v}$ & 30.46 & 3.66 & \multirow{2}{*}{2.439M} \\
{} & {} & $\mathbf{A}_\text{m}$ & 30.73 & 5.93 & {} \\\cline{2-6}
{} & \multirow{2}{*}{800} & $\mathbf{A}_\text{v}$ & \bf{32.28} & 4.39 & \multirow{2}{*}{6.478M} \\
{} & {} & $\mathbf{A}_\text{m}$ & 32.11 & 6.28 & {} \\\cline{2-6}
{} & \multirow{2}{*}{1600} & $\mathbf{A}_\text{v}$ & 31.94 & 4.68 & \multirow{2}{*}{19.356M} \\ 
{} & {} & $\mathbf{A}_\text{m}$ & 31.54 & 6.68 & {} \\ 
\hline
{\small\texttt{STFT/iSTFT}} & 1025 & N/A & N/A & \bf{8.80} &  N/A 
\end{tabular}}
\label{tab:res-2}
\vspace{-0.5cm}
\end{table}

\section{Conclusions}\label{sec:conclusions}
\vspace{-0.5cm}
In this work we presented a method for learning music signal representations in an unsupervised way. Our method is based on the denoising autoencoder model~\cite{vincent_den} and the differential digital signal processing concept~\cite{ddsp}. The befits of our method are interpretability, non-negativity for real-valued music signal representations for driving an established synthesis model, based on cosine functions. We conducted a series of experiments where we investigated the reconstruction capabilities of the proposed method subject to auto-encoding and informed source separation using binary masks. Our results demonstrate a reconstruction above $30$ dB of scale-invariant signal-to-distortion ratio, and that separation by masking is possible using the obtained representation. The latter, opens up directions for supervised approaches to masking-based separation. However, compared to the short-time Fourier transform and its inverse counterpart our results suggest that there is much room for improvements in order to achieve the benefits of the STFT. 

\vspace{-0.25cm}
\section*{Acknowledgements}
\vspace{-0.25cm}
Stylianos I. Mimilakis is supported in part by the German Reseach Foundation (AB 675/2-1, MU 2686/11-1).

\end{document}